\documentclass[%
 aip,
 amsmath,amssymb,
 reprint,%
]{revtex4-1}

\usepackage{graphicx}
\usepackage{dcolumn}
\usepackage{bm}
\usepackage{nidanfloat}

\usepackage{makecell}
\usepackage[utf8]{inputenc}
\usepackage[T1]{fontenc}
\usepackage{mathptmx}
\usepackage{etoolbox}
\usepackage{textgreek}
\usepackage{tabularx}
\usepackage{threeparttable}
\usepackage{placeins} 
\usepackage{pifont}
\makeatletter
\def\@email#1#2{%
 \endgroup
 \patchcmd{\titleblock@produce}
  {\frontmatter@RRAPformat}
  {\frontmatter@RRAPformat{\produce@RRAP{*#1\href{mailto:#2}{#2}}}\frontmatter@RRAPformat}
  {}{}
}
\makeatother
\begin{document}

\preprint{AIP/123-QED}

\title[Ultrahigh-Q chalcogenide micro-racetrack resonators]{Ultrahigh-Q chalcogenide micro-racetrack resonators}

\author{Bright Lu}
\affiliation{ 
Department of Electrical, Computer, and Energy Engineering, University of Colorado Boulder, 425 UCB, Boulder, Colorado 80309, USA}
\author{James W. Erikson}
\author{Bo Xu}
\affiliation{%
Department of Physics, University of Colorado Boulder, 390 UCB, Boulder, Colorado 80309, USA
}%
\author{Sinica Guo}
\affiliation{ 
Department of Electrical, Computer, and Energy Engineering, University of Colorado Boulder, 425 UCB, Boulder, Colorado 80309, USA
}
\author{Mo Zohrabi}
\affiliation{%
Department of Physics, University of Colorado Boulder, 390 UCB, Boulder, Colorado 80309, USA
}
\author{Juliet T. Gopinath}
\affiliation{ 
Department of Electrical, Computer, and Energy Engineering, University of Colorado Boulder, 425 UCB, Boulder, Colorado 80309, USA
}
\affiliation{%
Department of Physics, University of Colorado Boulder, 390 UCB, Boulder, Colorado 80309, USA
}
\affiliation{ 
Materials Science and Engineering Program, University of Colorado Boulder, 613 UCB, Boulder, Colorado 80309, USA
}

\author{Wounjhang Park}
\email{won.park@colorado.edu}
\affiliation{ 
Department of Electrical, Computer, and Energy Engineering, University of Colorado Boulder, 425 UCB, Boulder, Colorado 80309, USA
}
\affiliation{ 
Materials Science and Engineering Program, University of Colorado Boulder, 613 UCB, Boulder, Colorado 80309, USA
}
\date{\today}

\begin{abstract}
High-quality factor microresonators are an attractive platform for the study of nonlinear photonics, with diverse applications in communications, sensing, and quantum metrology. The characterization of loss mechanisms and nonlinear properties in a microresonator is a necessity for the development of photonic integrated circuits. Here, we demonstrate a high-quality chalcogenide (Ge\textsubscript{23}Sb\textsubscript{7}S\textsubscript{70}) micro-racetrack resonator utilizing Euler curves. The racetrack geometry is studied to minimize loss at both the straight--curved waveguide junction and through the waveguide curve. The material absorption, intrinsic quality factor, and nonlinear index are extracted by a comprehensive model fit to laser wavelength resonance scans. The micro-racetrack resonator possesses an absorption loss of 0.43 dB/m, an intrinsic quality factor of 4.5 $\times$ 10\textsuperscript{6}, and nonlinear index of 1.28 $\times$ 10\textsuperscript{-18} m\textsuperscript{2}/W, in a waveguide cross-section less than 1 {\textmu}m\textsuperscript{2}. Our results yield state-of-the-art nonlinear microresonators and establish Ge\textsubscript{23}Sb\textsubscript{7}S\textsubscript{70} as a low-loss PIC platform. 


\vspace{12mm}
\end{abstract}

\maketitle

\section{\label{sec:level1}Introduction}
\vspace{0mm}

High-quality optical microresonators provide efficient access to nonlinear light-matter interactions by resonant optical enhancement and are an essential component of photonic integrated circuits (PICs). Advances have spawned diverse on-chip applications, including telecommunications, metrology, and quantum information processing \cite{spencer18synth, kippcomb, wang_integrated_2020}. However, PICs require low propagation losses and high effective nonlinearity to surpass the milliwatt-level efficiency of off-chip solutions\cite{liu_high-yield_2021,zhou_prospects_2023}. The development of low-loss, highly nonlinear PICs is therefore of significant interest for the commercialization of on-chip technologies: hence, low-loss microresonators have been investigated in a variety of material candidates, including Group IV semiconductors, III-V semiconductors, and dielectrics \cite{Siliconphotonicsring, zhang_mid-infrared_2022, Pfeiffer:16, Zhu:21}, each possessing unique optical properties and processing considerations.

Among PIC material candidates, chalcogenides are promising for their large nonlinear refractive indices (1 -- 10 $\times$ 10\textsuperscript{-18} m\textsuperscript{2}/W) and wide infrared transparency (up to 20 {\textmu}m) \cite{eggleton2011chalcogenide}. Their amorphous nature enables the thermal evaporation of chalcogenides on a wide variety of substrates, circumventing bonding processes and enabling unparalleled flexibility in microresonator design \cite{grayson2022fabrication}. This combination of properties has allowed for the demonstration of near- and mid-infrared nonlinear phenomena in chalcogenides, including Brillouin lasing, supercontinuum generation, and Kerr soliton formation \cite{Kabakova:13, ko2025mid, Li24Brillouin, grayson2019enhancement, xia2023integrated}. Still, chalcogenide microresonator losses remain high (> 10 dB/m) relative to drawn fiber \cite{shiryaev_preparation_2009}. Planar chalcogenide microresonators have thus been limited to quality factors of 0.5 – 5 × 10\textsuperscript{6}, which necessitates milliwatt-level pump power for nonlinear interactions and hinders their widespread adoption \cite{Li24Brillouin, xia2023integrated, Zhang21As2S3, wang2023high}.

The majority of propagation loss in a microresonator can be attributed to roughness-induced scattering at the waveguide sidewall and higher-order mode excitation through the waveguide curve\cite{siliconracetrack,kitoh1992}. Increasing the waveguide width and bend radius will mitigate these losses, at the cost of integration density. An effective alternative is to adopt waveguide curves based on Euler spirals, where the curvature changes linearly with arc length\cite{siliconracetrack, kang2025high}. Euler curves have been demonstrated to suppress the excitation of higher-order modes, enabling lower microresonator losses with small device footprints.

The remainder of microresonator propagation loss lies in absorption at the waveguide surface and within the guiding material. Absorption loss is typically small (< 5 dB/m), and is therefore difficult to extract from microresonator propagation loss measurements dominated by scattering and bending losses. The characterization of absorption loss is particularly significant in chalcogenide-based PICs, which often suffer from non-negligible absorption losses due to surface oxidation and impurity-escalated material absorption\cite{Zhang21As2S3, ko2025mid}. 

In this work, we present an investigation of propagation loss mechanisms in a chalcogenide micro-racetrack resonator. We reveal that the micro-racetrack resonator loss stems from both higher-order mode excitation through the waveguide curve and fundamental mode mismatching loss at the straight--curved waveguide junction, which can be effectively mitigated by selecting an appropriate waveguide width and minimum radius. Further, we use a comprehensive model based on coupled-mode theory\cite{zhu2019nonlinear} to directly extract the material absorption and nonlinear index from resonance transmission spectra. The chalcogenide microresonator exhibits an absorption loss of 0.43 dB/m, an intrinsic quality factor of 4.55 $\times$ 10\textsuperscript{6}, and nonlinear index of 1.28 $\times$ 10\textsuperscript{-18} m\textsuperscript{2}/W, in a waveguide cross-section less than 1 {\textmu}m\textsuperscript{2}. 
This work represents the highest nonlinear figure of merit, as defined by the ratio of effective nonlinearity and loss, reported in chalcogenide-based PICs\cite{Zhang21As2S3}.

\break

\begin{figure}[!htbp]
  \vspace{1mm}
  \includegraphics[width=\linewidth]{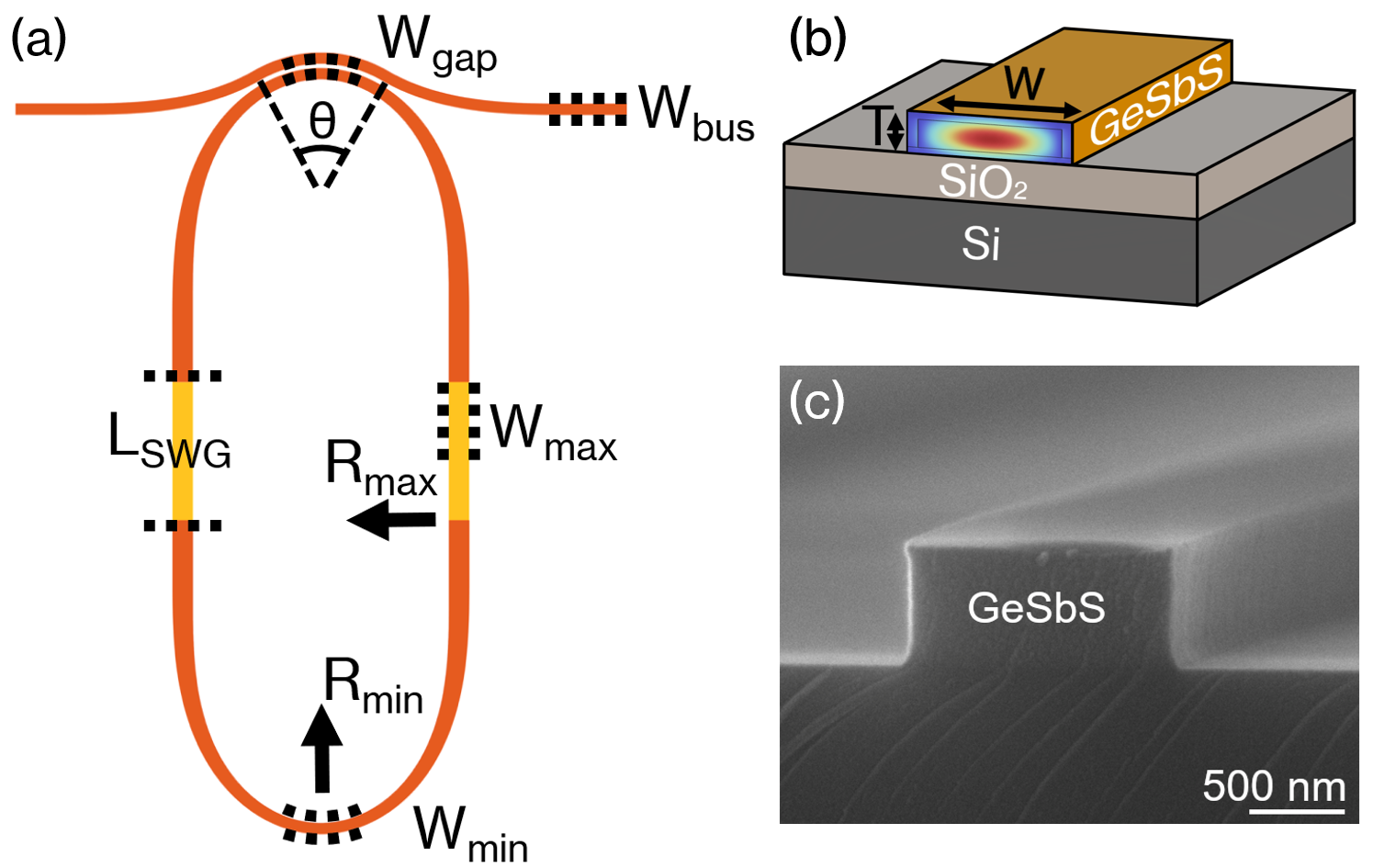}
  \caption{(a) Proposed micro-racetrack resonator schematic. (b) Chalcogenide waveguide cross-section. The inset shows the simulated fundamental TE mode in a representative waveguide. (c) Scanning electron micrograph of a fabricated 0.6 $\times$ 1.4 {\textmu}m chalcogenide waveguide cross-section.}
  \label{fig:boat1}
\end{figure}

\vspace{-10mm}
\section{\label{sec:level2}Microresonator design}

Figure 1(a) shows the schematic of the proposed micro-racetrack resonator. The racetrack design connects straight waveguide (SWG) sections of width $W\textsubscript{max}$ and length $L\textsubscript{SWG}$ to tapered Euler curves with minimum width $W\textsubscript{min}$ and effective curvature described by radii $R\textsubscript{max}$ and $R\textsubscript{min}$. The taper eases minimum dimensions imposed on the coupling gap $W\textsubscript{gap}$ and waveguide width $W\textsubscript{bus}$ of the directional pulley coupler by leaking the evanescent field of the racetrack fundamental mode to assist coupling, thus improving design robustness in fabrication. The pulley coupler, with angle $\theta$, is implemented to minimize coupling to higher-order modes and further relax minimum required dimensions of $W\textsubscript{gap}$ and $W\textsubscript{bus}$. 

Euler curves are implemented to reduce radiative bend loss and higher-order mode excitation through the waveguide curve while preserving device footprint\cite{siliconracetrack, kang2025high}. The Euler curve linearly decreases its curvature along its path length, from maximum curvature $R\textsubscript{max}$ to minimum curvature $R\textsubscript{min}$, as defined by \cite{Jiang:18}
\begin{eqnarray}
\frac{d\theta}{dL}=\frac{1}{R}=\frac{L}{A^2}+\frac{1}{R_{max}}
\label{eq:one}.
\end{eqnarray}

\noindent where $A$ is a constant, defined by $A=[L_{total}/(1/R_{min}-1/R_{max})]^{1/2}$, and $L\textsubscript{total}$ is the length of the Euler curve. Two 90$^\circ$ Euler curves are combined to make the 180$^\circ$ Euler curve implemented in the racetrack. 

The waveguide core consists of $T$ = 500 nm thick Ge\textsubscript{23}Sb\textsubscript{7}S\textsubscript{70} chalcogenide glass, deposited on top of a 2 {\textmu}m thick thermal oxide on silicon wafer. Figures 1(b) and 1(c) show the proposed and fabricated chalcogenide waveguide cross-section. The microresonator is designed for the infrared C- and L-band, and TE operation is selected to minimize mode overlap with the waveguide sidewall. The refractive index is measured to be $n_{GeSbS}$ = 2.196 at $\lambda$ = 1550 nm.

The propagating mode in a perfect micro-racetrack resonator (i.e., negligible material absorption, sidewall-scattering) has four loss mechanisms: fundamental mode mismatching loss at the straight--curved waveguide junction, higher-order mode excitation at the junction, higher-order mode excitation through the waveguide curve, and radiative bend loss through the curve. For large $R$ (> 50 {\textmu}m) and wide $W$ (> $\lambda$) waveguide curves, radiative bend loss is negligible\cite{Jiang:18}. 

Fundamental mode mismatching and higher-order mode excitation loss at the junction can be significantly mitigated by selecting appropriate $W\textsubscript{max}$ and $R\textsubscript{max}$. To identify a local minimum of straight--curved waveguide junction loss, mode field distributions of TE\textsubscript{0} and TE\textsubscript{1} for straight and curved waveguides of varying width and radii are calculated (COMSOL Multiphysics). We then calculate overlap integrals between the TE\textsubscript{0} straight waveguide mode and the TE\textsubscript{0,1} curved waveguide modes to predict the mode mismatch and higher-order mode excitation loss at the junction.

\begin{figure}[!htbp]
    \vspace{-3mm}
  \includegraphics[width=\linewidth]{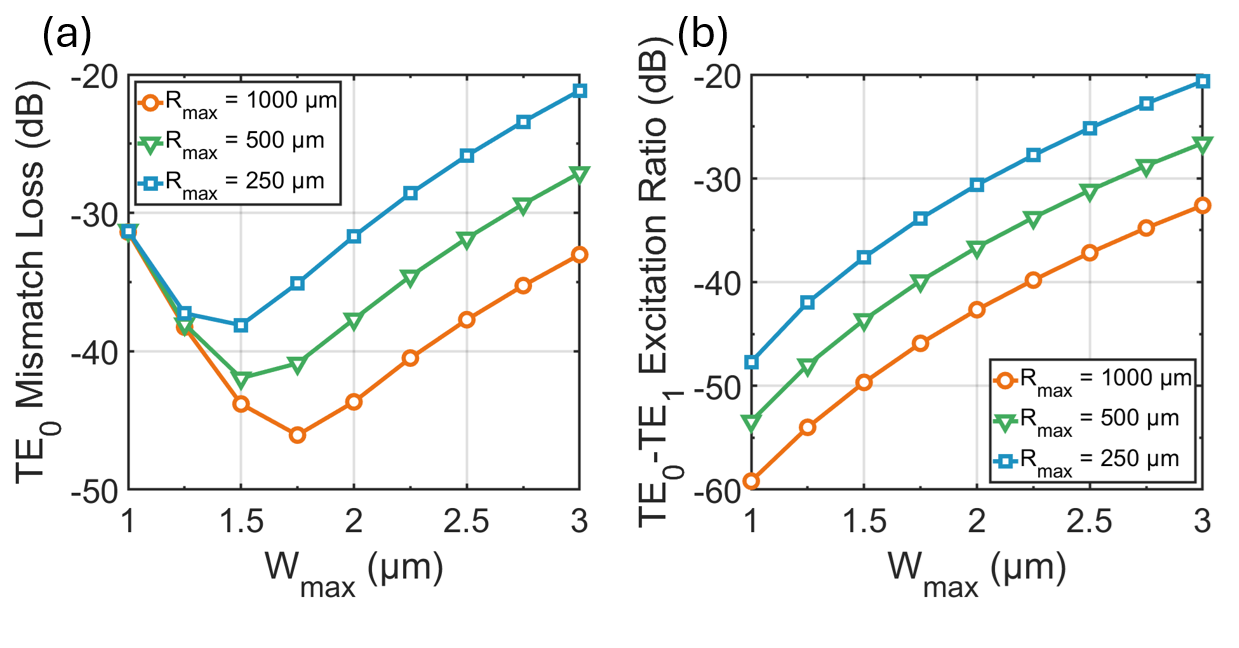}
  \vspace{-8mm}
  \caption{(a) Calculated mode mismatch loss of the fundamental TE mode as it enters Euler curves of varied maximum radii $R\textsubscript{max}$. (b) Calculated excitation of the TE\textsubscript{1} mode as the TE\textsubscript{0} mode enters Euler curves of varied $R\textsubscript{max}$.}
  \label{fig:fig2}
  \vspace{-3mm}
\end{figure}

Figures 2(a) and (b) show the TE\textsubscript{0} mode mismatch and higher-order mode excitation junction loss dependence on $W\textsubscript{max}$ and $R\textsubscript{max}$ for a 500 nm thick GeSbS waveguide. We observe that TE\textsubscript{0} loss from both mechanisms can be mitigated by selecting a larger $R\textsubscript{max}$. Further, a local minimum of TE\textsubscript{0} mismatch loss is found near $W\textsubscript{max}$ = 1.75 {\textmu}m for $R\textsubscript{max}$ = 1000 {\textmu}m. This minimum mismatch loss can be intuitively understood by the slight displacement of the fundamental mode in the waveguide curve from the waveguide center. As the waveguide width increases, the fundamental mode first becomes more tightly confined to the waveguide core, then begins to shift towards the outer edge. Implementing a lateral offset to the straight waveguide towards the outer edge of the racetrack would reduce mismatch loss further. Thus, we selected $R\textsubscript{max}$ = 1000 {\textmu}m and $W\textsubscript{max}$ = 1.9 {\textmu}m to minimize total junction loss, while maintaining a large waveguide width to inhibit sidewall scattering loss.

Loss stemming from higher-order mode excitation through the waveguide curve can be effectively negated by selecting an adequately large $R\textsubscript{min}$. Figures 3(a-c) show simulated higher-order TE excitation from the TE\textsubscript{0} mode (FDTD, Lumerical MODE Solutions) through 180$^\circ$ Euler curves of varied minimum curvature $R\textsubscript{min}$ with fixed $R\textsubscript{max}$ = 1000 {\textmu}m, $W\textsubscript{max}$ = 1.9 {\textmu}m, and $W\textsubscript{min}$ = 1.5 {\textmu}m. $W\textsubscript{max}$ and $R\textsubscript{max}$ are selected to minimize total junction loss and $W\textsubscript{min}$ is selected to taper the waveguide curve, thus aiding the directional coupler. As $R\textsubscript{min}$ is increased from 20 {\textmu}m to 50 {\textmu}m, the TE\textsubscript{0} excitation of TE\textsubscript{1} and TE\textsubscript{2} declines. Figure 3(d) shows the calculated\break

\vspace{-8mm}
\begin{figure}[!htbp]
  \includegraphics[width=\linewidth]{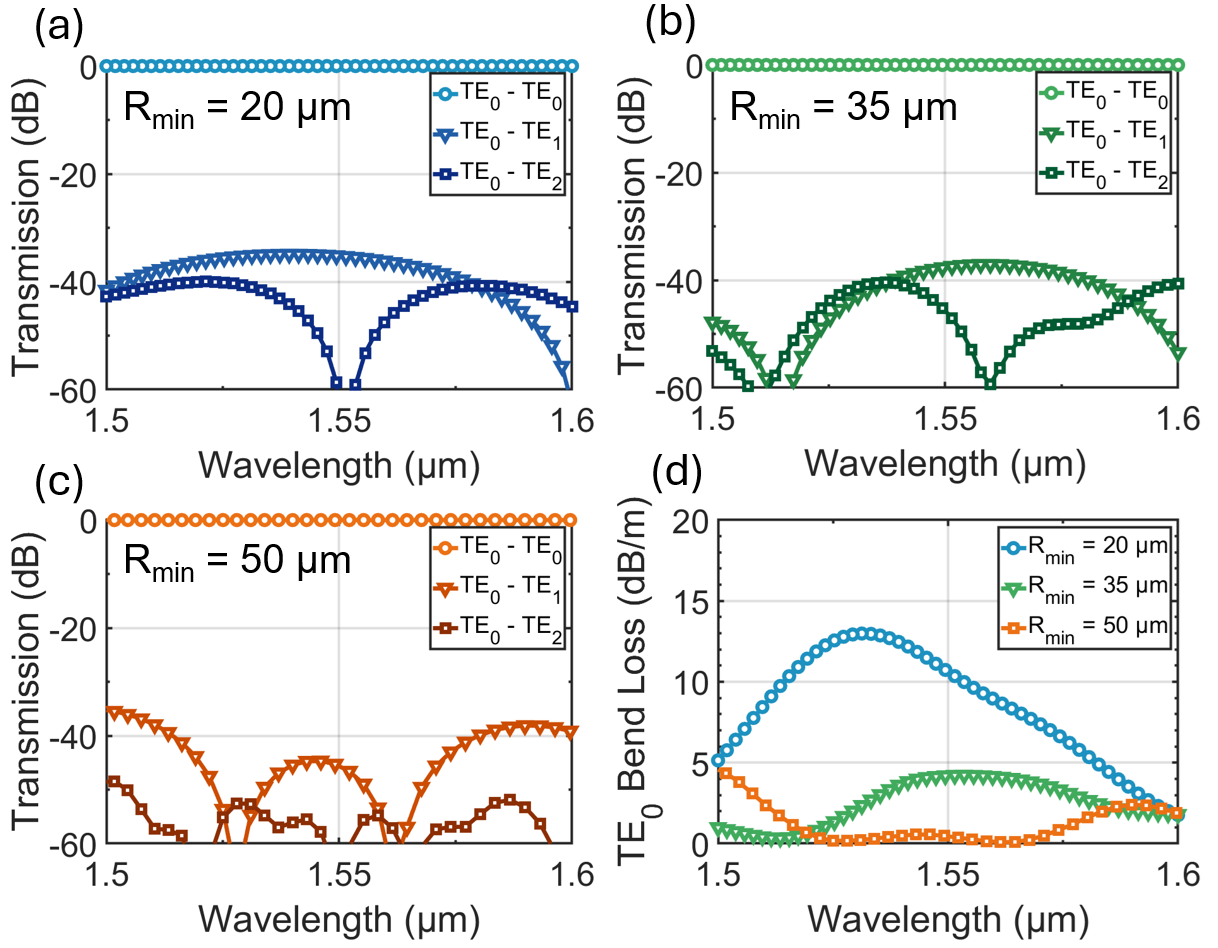}
  \vspace{-6mm}
  \caption{Calculated light propagation through a 180$^\circ$ Euler curve with $R\textsubscript{max}$ = 1000 {\textmu}m and (a) $R\textsubscript{min}$ = 20 {\textmu}m, (b) $R\textsubscript{min}$ = 35 {\textmu}m, and (c) $R\textsubscript{min}$ = 50 {\textmu}m. (d) Corresponding bend loss of the fundamental TE mode for varied $R\textsubscript{min}$.}
  \label{fig:fig3}
      \vspace{-3mm}
\end{figure}

\noindent bend loss of the TE\textsubscript{0} mode monitored at the end of the Euler curve, taking into account loss stemming from excited higher-order modes and potential radiative loss through the curve. For $R\textsubscript{min}$ = 50 {\textmu}m, total predicted loss through the Euler curve is below 1 dB/m. Therefore, an $R\textsubscript{max}$ = 1000 {\textmu}m, $R\textsubscript{min}$ = 50 {\textmu}m, $W\textsubscript{max}$ = 1.9 {\textmu}m, and $W\textsubscript{min}$ = 1.5 {\textmu}m were selected for the final design. To avoid stitching errors during electron beam lithography, $L\textsubscript{SWG}$ was selected such that the device footprint, including directional coupler, fit within a single write field.

\vspace{-4mm}
\section{\label{sec:level3}Experiments}

The chalcogenide micro-racetrack resonator is fabricated as follows. First, 500 nm of Ge\textsubscript{23}Sb\textsubscript{7}S\textsubscript{70} is thermally evaporated onto a 2 {\textmu}m thick thermal oxide on silicon wafer. Ge\textsubscript{23}Sb\textsubscript{7}S\textsubscript{70} was selected for its non-toxicity, high linear and nonlinear index, and long-term stability. The chalcogenide thin film is then annealed in vacuum at 250$^\circ$C, below glass transition temperature, for two hours to reduce surface roughness and improve material homogeneity. Next, the wafer is coated with 300 nm of AR-P 6200 e-beam resist and patterned using direct-write 100 kV electron beam lithography (Raith EBPG5200ES). The resist is then developed in room temperature AR 600-546 for 60s, and the device is etched in an inductively coupled plasma reactive ion etching chamber using a mixture of BCl\textsubscript{3} and Ar gases. The remaining resist layer is stripped in room temperature N-methyl-2-pyrrolidone for 16 hours and fully removed following a 60 s ultrasonic bath. 

Optical resonances are probed using a mode-hop-free, tunable narrow-linewidth laser in the 1550 nm band (Toptica CTL 1550). Light is coupled into the waveguides using a free-space alignment procedure, making use of aspheric lenses (Thorlabs C660TME-C) to focus the incoming beam onto the waveguide facet. This allows for precise, free-space control of laser polarization as well as direct measurement of input laser power while reducing coupling losses to approximately 4 dB per facet. Light from the waveguide output facet is similarly collected with an aspheric lens and focused onto a high-speed, biased InGaAs photodetector (Thorlabs DET10C).

Once optical coupling has been achieved, the wavelength is swept over a wide range in order to probe the resonator free spectral range, identify supported optical modes, and ensure input polarization is as expected. This wide scan also allows modally pure, high quality resonances to be identified for further investigation. The resonances observed during this wide scan are initially fitted using a Lorentzian model to determine the loaded Q factor ($Q_{l}$). The intrinsic Q factor ($Q_{int}$) can then be estimated using $Q_{int}= 2Q_{l} / (1 \pm \sqrt{T_0})$, where $T_0$ is the minimum normalized resonance transmission\cite{zhang_mid-infrared_2022}. The + or - is determined by under- or over-coupling respectively.

\begin{figure}[!htbp]
  \includegraphics[width=\linewidth]{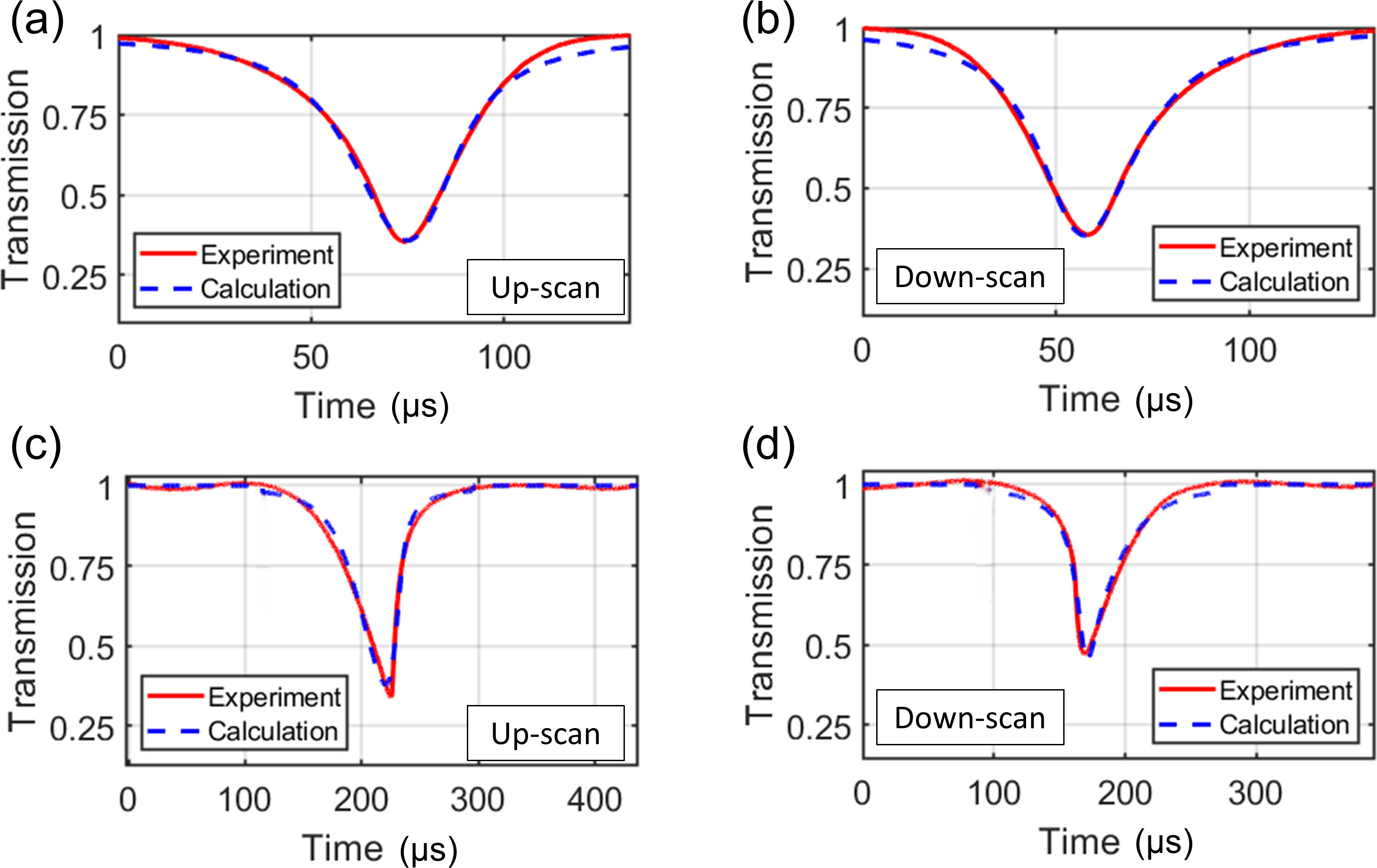}
  \vspace{-6mm}
  \caption{Experimental transmission spectra for scans in both the red (up) and blue (down) directions at two different input powers. (a) Low power up-scan (b) Low power down-scan (c) High power up-scan (d) High power down-scan. Spectra are fit simultaneously using a genetic algorithm to prevent overfitting and give more accurate fitting parameters.}
  \label{fig:ResFits}
\end{figure}
\vspace{-2mm}
To further characterize the device properties, we employ a nonlinear model developed by Zhu \textit{et al.}\cite{zhu2019nonlinear} based on coupled-mode theory. This model not only provides an accurate measurement of $Q_{int}$, which can be used to validate the Lorentzian method, but also allows for the direct extraction of nonlinear and thermal properties, which are difficult to isolate in fabricated microstructures. The nonlinear model employs the simultaneous solution of a set of seven coupled equations and accepts six fitting parameters\cite{zhu2019nonlinear}: intrinsic and coupling Q factors ($Q_{int}, Q_{c}$), nonlinear index ($n_2$), optical absorption coefficient ($\alpha$), thermal constant ($\tau$), and optical backcoupling rate ($g$). To collect data for the nonlinear model, high $Q_{l}$ resonances are selected and probed using a narrow wavelength scan performed by piezo-actuating the laser cavity, scanning the laser frequency at a rate of 8 GHz/s. Data is recorded while scanning both in the red and blue directions at multiple laser powers ranging from 0.5 mW to 2 mW. Multiple scans are fit simultaneously to reduce overfitting, using a genetic algorithm implemented in MATLAB. Results of the fitting process can be seen in Figure \ref{fig:ResFits}.

The quality of the resulting fits is assessed using the mean squared error, which is minimized by the genetic algorithm. The main source of error is the presence of an oscillatory background in the data. This is caused by back-reflections\break

\vspace{-8mm}
\begin{table}[!htbp]
\centering
\begin{ruledtabular}
\begin{tabular}{l r} 
 
 \textbf{Parameter} & \textbf{Value} \\ [0.5ex] 
 \hline
 $Q_{int}$ (10\textsuperscript{6}) & 4.55 \\ 
 $n_2$ (10\textsuperscript{-18} m\textsuperscript{2}/W) & 1.28 \\
 $\alpha_{abs}$ (10\textsuperscript{-4}/m) & 983 \\
 $\tau$ ({\textmu}s) & 2.50 \\
 $g$ (MHz) & 189.4 \\
\end{tabular} 
\end{ruledtabular}
\vspace{-2mm}
\caption{Best-fit parameters for simultaneous fitting up- and down-scan spectra across two different input powers. Values fall within expected ranges for the resonator material and geometry.}
\label{table:fit_parameters}
\vspace{-4mm}
\end{table}
\vspace{-1.5mm}

\noindent from the flat facets at the ends of the bus waveguide. These reflections produce a weak Fabry-Perot cavity effect, causing the transmission spectrum to exhibit a sinusoidal background. To account for this, we apply a polynomial fit to the background transmission around each resonance. The data is then normalized using this fit as a baseline, resulting in a cleaner representation of the resonance and significant improvement in the nonlinear fits.

The best-fit parameters returned by the genetic algorithm are shown in Table \ref{table:fit_parameters}. In each case, the results fall within the expected ranges for the material platform. We report a measured $Q_{int}$ value of $4.55\times10^6$, which agrees with the Lorentzian estimation process described above. The nonlinear index is found to be $1.28\times10^{-18}$ m\textsuperscript{2}/W, which, while being nearly fifty times larger than silica, is slightly lower than the bulk nonlinearities of Ge-Sb-S based glasses reported from Z-scan measurements\cite{Hu:21} and self-phase modulation experiments\cite{choi2016nonlinear} taken near 1550 nm, yielding values ranging from $1.7$ to $3.71\times10^{-18}$ m\textsuperscript{2}/W. The difference likely stems from thermal evaporation, which alters the relative stoichiometry of Ge-Sb-S from the bulk material\cite{grayson2019enhancement, li_study_2022}. The measured value of $\alpha_{abs}$ is 0.098/m, which corresponds to an absorption loss of 0.43 dB/m. This absorption loss is comparable to minimum propagation losses measured in drawn chalcogenide fiber\cite{shiryaev_preparation_2009, Parnell:19, Hu:21}, indicating that our thin film is nearly impurity-free. We measure a value of 2.5  {\textmu}s for $\tau$, representing a fast cooling process inherent to the planar waveguide structure, where heat can be easily dissipated into the substrate. Finally, we measure a large  backscattering parameter, which is likely dominated by back-reflections from the output bus waveguide facet.

\begin{figure}[!htbp]
    \vspace{-3mm}
  \includegraphics[width=\linewidth]{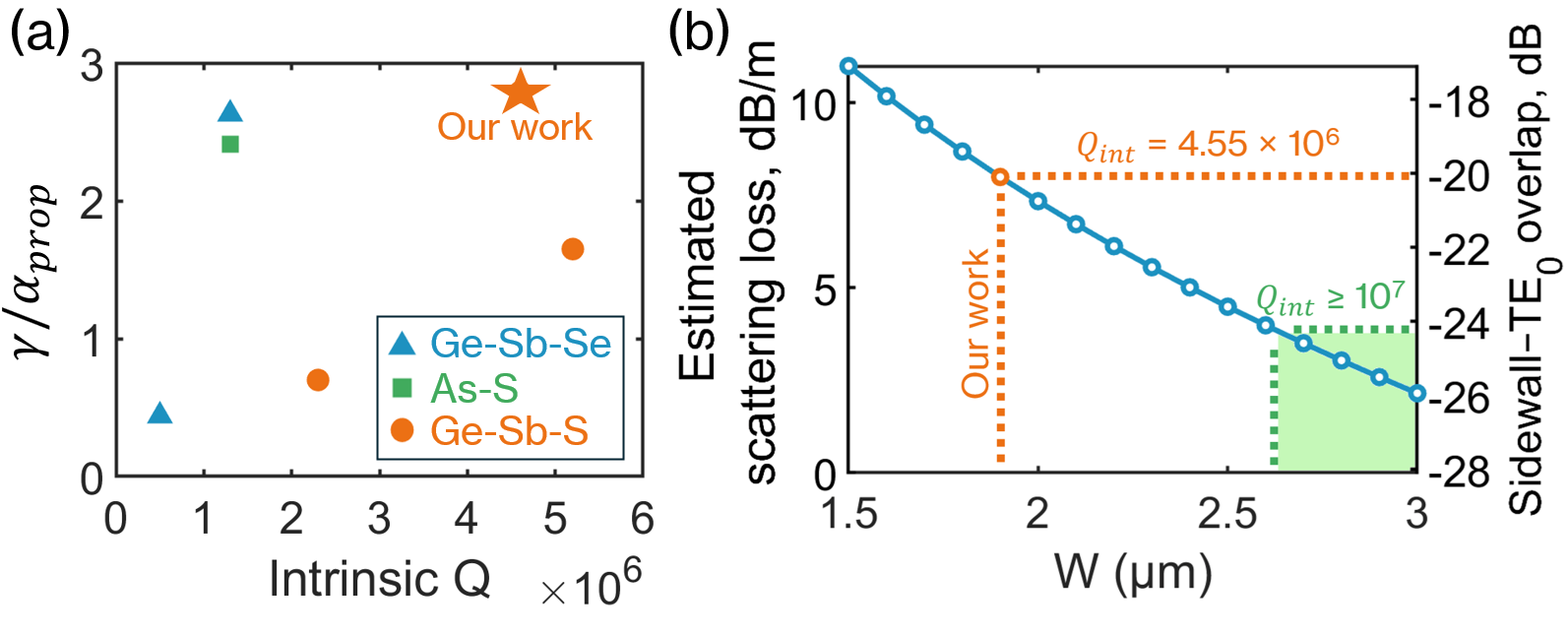}
  \vspace{-8mm}
  \caption{(a) Effective nonlinear performance of high-Q chalcogenide-based PICs. (b) Estimated micro-racetrack scattering loss mitigation using wider waveguide geometries.}
  \label{fig:expectedQ}
\end{figure}

Table II compares the demonstrated performance of planar chalcogenide microresonators for integrated nonlinear photonics near 1550 nm. Due to their large nonlinear indices, Ge\textsubscript{28}Sb\textsubscript{12}Se\textsubscript{60}- and As\textsubscript{2}S\textsubscript{3}-based devices have been selected for comparison. To compare the effective mode area $A_{eff}$, material nonlinearity $n_2$, and propagation loss $\alpha_{prop}$ between chalcogenide devices, we apply a nonlinear figure of merit, defined by $\gamma$/$\alpha_{prop}$\cite{Zhang21As2S3}, where the effective nonlinear coefficient $\gamma=2\pi n_2/(\lambda A_{eff})$. A large figure of merit indicates a highly nonlinear, low-loss device. Our proof-of-concept device possesses a small waveguide cross-section and low propagation loss, yielding the highest achieved figure of merit in chalcogenide planar microresonators, as shown in Figure \ref{fig:expectedQ}(a). We believe that our improved performance is due to our racetrack design: while other high-Q chalcogenide microresonators have implemented Euler curves, our design is the first, to the best of our knowledge, to investigate in detail the loss contributions at the straight--curved waveguide junction. In addition, our design may be implemented in tandem with other reported loss mitigation methods in chalcogenides, such as purification treatments\cite{ko2025mid}, annealings\cite{Zhang21As2S3}, and optical claddings\cite{xia2023integrated,Li24Brillouin}, for higher-efficiency nonlinear PICs. 

\vspace{-4mm}

Our current device possesses propagation loss totaling 8.5 dB/m. We predict that, by widening the waveguide, our micro-racetrack design may be readily extended to achieve nearly impurity-limited performance. From our fitted absorption loss of 0.43 dB/m, assuming remaining bend loss to be negligible, we expect our propagation loss to be dominated by optical scattering. Scattering occurs at the waveguide sidewall, and can therefore be mitigated by reducing sidewall overlap with the TE\textsubscript{0} mode. Correlating sidewall--TE\textsubscript{0} overlap with scattering loss, we find that we can reduce scattering loss by more than 4 dB, using waveguide widths exceeding 2.6 {\textmu}m, as shown in Figure \ref{fig:expectedQ}(b). Such dimensions are comparable to other high-Q resonators reported in literature\cite{xia2023integrated,wang2023high,kang2025high}, although care must be taken to mitigate mode mismatch loss at the straight--curved waveguide junction\cite{irfan_ultra-compact_2024,kitoh1992}. 










\onecolumngrid
\begin{table*}[!htbp]
\vspace{-4mm}
\begin{minipage}{\textwidth}
\centering
\begin{threeparttable}[!htbp]
\caption{Comparison of effective nonlinearity, propagation loss, and intrinsic Q factor in planar chalcogenide microresonators.}
\renewcommand*{\arraystretch}{1.3}
\noindent
\newcolumntype{b}{>{\hsize=1.2\hsize}l}
\newcolumntype{s}{>{\hsize=.5\hsize}l}
\newcolumntype{m}{>{\hsize=0.8\hsize}l}
\begin{ruledtabular}
\begin{tabular}{m m s s b m m s s s s s m} 
 \textbf{Ref.} & \textbf{Material} & \textbf{\textit{n}} & \makecell[l]{\textbf{\textit{n\textsubscript{2}}} \\ \textbf{(m\textsuperscript{2}/W)}}  & \textbf{Geometry} & \textbf{Cladding} & \makecell[l]{\textbf{Dimensions} \\  \textbf{({\textmu}m\textsuperscript{2})}} & \makecell[l]{\textbf{$\gamma$} \\  \textbf{(m\textsuperscript{-1}W\textsuperscript{-1})}} & \textbf{\textit{R} ({\textmu}m)} & \makecell[l]{\textbf{$\alpha_{prop}$} \\ \textbf{(dB/m)}} & \makecell[l]{\textbf{FOM} \\ \textbf{($\gamma$/$\alpha$)}} & \makecell[l]{\textbf{\textit{Q\textsubscript{int}}} \\ \textbf{($\times$ 10\textsuperscript{6})}}\\ 
 \hline
 {[\!\citenum{grayson2022fabrication}]} & Ge-Sb-Se & N/A & 5.1\tnote{a} & Micro-ring & Air-clad & 0.4 $\times$ 0.9 & 14.6\tnote{b} & 20 & 144 & 0.44\tnote{b} & 0.4 \\ [-.1em]
{[\!\citenum{wang2023high}]} & Ge-Sb-Se & 2.77 & 5.1\tnote{a} & Micro-racetrack & SU-8 & 0.3 $\times$ 3.0 & 23.0\tnote{b} & 65 & 38 & 2.63\tnote{b} & 1.3 \\ [-.1em]
{[\!\citenum{Zhang21As2S3}]} & As-S & 2.43 & 3 & Micro-ring & BCB & 0.85 $\times$ 2.0 & 10.0 & 100 & 18 & 2.41 & 1.3 \\ [-.1em]
{[\!\citenum{xia2023integrated}]} & Ge-Sb-S & 2.2 & 1.3 & Micro-ring & SiO\textsubscript{2} & 0.8 $\times$ 2.4 & 2.74\tnote{b} & 100 & 17\tnote{b} & 0.70\tnote{b} & 2.3 \\ [-.1em]
{[\!\citenum{Li24Brillouin}]} & Ge-Sb-S & 2.2\tnote{a} & 1.3\tnote{a} & Micro-racetrack & SiO\textsubscript{2} & 0.85 $\times$ 2.4 & 2.58\tnote{b} & 100 & 6.8\tnote{b} & 1.65\tnote{b} & 5.2 \\ [-.1em]
\textbf{This work} & Ge-Sb-S & \textbf{2.19} & \textbf{1.28} & \textbf{Micro-racetrack} & \textbf{Air-clad} & \textbf{0.5 $\times$ 1.9} & \textbf{5.46} & \textbf{100} & \textbf{8.5} & \textbf{2.79} & \textbf{4.6} \\ [-.1em]

\end{tabular}
\end{ruledtabular}
\begin{tablenotes}
\item[a] Representative values taken from literature.
\item[b] Calculated values from reported data and literature.
\end{tablenotes}
\end{threeparttable}
\label{table:lit_comparison}
\vspace{-3mm}
\end{minipage}
\end{table*}
\twocolumngrid

\vspace{-3mm}
\section{\label{sec:level4}Conclusion}

Photonic integrated circuits (PICs) are a power-efficient, scalable solution for next-generation telecommunications and information processing. High-quality optical microresonators are necessary to efficiently access optical nonlinearities and surpass off-chip optical systems. Our design yields state-of-the-art nonlinear microresonators by nullifying bend losses, enabled through a comprehensive investigation of propagation loss mechanisms in racetrack geometries. 

High-quality chalcogenide micro-racetrack resonators were demonstrated by incorporating Euler curves. The waveguide geometry design space was explored to minimize higher-order mode excitation and modal mismatch loss through the racetrack, enabling compact microresonators with significantly reduced round-trip loss. Further, we applied a coupled-mode theory model to directly extract the intrinsic quality factor, nonlinear index, and absorption. The fabricated micro-racetrack resonator has an intrinsic quality factor of 4.55 $\times$ 10\textsuperscript{6} and absorption loss of just 0.43 dB/m. The high-Q performance immediately enables the efficient access of stimulated Brillouin and Raman scattering for narrow linewidth lasers. The low-loss micro-racetrack design is also compatible with dispersion engineering methods and is therefore a promising platform for four-wave mixing-based applications.

\vspace{-3mm}
\section*{Acknowledgments}

This work was funded by the National Science Foundation (NSF ECCS 2224065). The authors thank Jiangang Zhu for his assistance in developing the coupled-mode theory model. The authors also thank the University of Delaware Nanofabrication Facility and the Colorado Shared Instrumentation for Nanofabrication and Characterization for their cleanroom facilities and expertise. 
\vspace{-3mm}
\section*{Conflict of interest}
\vspace{-0mm}
The authors declare no conflict of interest.
\vspace{-3mm}
\section*{Data Availability}
\vspace{0mm}
The data that support the findings of this study are available from the corresponding author upon reasonable request.
\vspace{-3mm}
\section*{References}
\vspace{-2mm}
\bibliographystyle{aapmrev4-1}
\bibliography{references}

\begin{thebibliography}{29}%
\makeatletter
\providecommand \@ifxundefined [1]{%
 \@ifx{#1\undefined}
}%
\providecommand \@ifnum [1]{%
 \ifnum #1\expandafter \@firstoftwo
 \else \expandafter \@secondoftwo
 \fi
}%
\providecommand \@ifx [1]{%
 \ifx #1\expandafter \@firstoftwo
 \else \expandafter \@secondoftwo
 \fi
}%
\providecommand \natexlab [1]{#1}%
\providecommand \enquote  [1]{``#1''}%
\providecommand \bibnamefont  [1]{#1}%
\providecommand \bibfnamefont [1]{#1}%
\providecommand \citenamefont [1]{#1}%
\providecommand \href@noop [0]{\@secondoftwo}%
\providecommand \href [0]{\begingroup \@sanitize@url \@href}%
\providecommand \@href[1]{\@@startlink{#1}\@@href}%
\providecommand \@@href[1]{\endgroup#1\@@endlink}%
\providecommand \@sanitize@url [0]{\catcode `\\12\catcode `\$12\catcode `\&12\catcode `\#12\catcode `\^12\catcode `\_12\catcode `\%12\relax}%
\providecommand \@@startlink[1]{}%
\providecommand \@@endlink[0]{}%
\providecommand \url  [0]{\begingroup\@sanitize@url \@url }%
\providecommand \@url [1]{\endgroup\@href {#1}{\urlprefix }}%
\providecommand \urlprefix  [0]{URL }%
\providecommand \Eprint [0]{\href }%
\providecommand \doibase [0]{http://dx.doi.org/}%
\providecommand \selectlanguage [0]{\@gobble}%
\providecommand \bibinfo  [0]{\@secondoftwo}%
\providecommand \bibfield  [0]{\@secondoftwo}%
\providecommand \translation [1]{[#1]}%
\providecommand \BibitemOpen [0]{}%
\providecommand \bibitemStop [0]{}%
\providecommand \bibitemNoStop [0]{.\EOS\space}%
\providecommand \EOS [0]{\spacefactor3000\relax}%
\providecommand \BibitemShut  [1]{\csname bibitem#1\endcsname}%
\let\auto@bib@innerbib\@empty
\bibitem [{\citenamefont {Spencer}\ \emph {et~al.}(2018)\citenamefont {Spencer}, \citenamefont {Drake}, \citenamefont {Briles}, \citenamefont {Stone}, \citenamefont {Sinclair}, \citenamefont {Frederick}, \citenamefont {Li}, \citenamefont {Westly}, \citenamefont {Ilic}, \citenamefont {Bluestone}, \citenamefont {Volet}, \citenamefont {Komljenovic}, \citenamefont {Hoon}, \citenamefont {Yoon}, \citenamefont {Suh}, \citenamefont {Youl}, \citenamefont {Pfeiffer}, \citenamefont {Kippenberg}, \citenamefont {Norberg}, \citenamefont {Vahala}, \citenamefont {Srinivasan}, \citenamefont {Newbury}, \citenamefont {Theogarajan}, \citenamefont {Bowers}, \citenamefont {Diddams},\ and\ \citenamefont {Papp}}]{spencer18synth}%
  \BibitemOpen
  \bibfield  {author} {\bibinfo {author} {\bibfnamefont {D.}~\bibnamefont {Spencer}}, \bibinfo {author} {\bibfnamefont {T.}~\bibnamefont {Drake}}, \bibinfo {author} {\bibfnamefont {T.}~\bibnamefont {Briles}}, \bibinfo {author} {\bibfnamefont {J.}~\bibnamefont {Stone}}, \bibinfo {author} {\bibfnamefont {L.}~\bibnamefont {Sinclair}}, \bibinfo {author} {\bibfnamefont {C.}~\bibnamefont {Frederick}}, \bibinfo {author} {\bibfnamefont {Q.}~\bibnamefont {Li}}, \bibinfo {author} {\bibfnamefont {D.}~\bibnamefont {Westly}}, \bibinfo {author} {\bibfnamefont {B.}~\bibnamefont {Ilic}}, \bibinfo {author} {\bibfnamefont {A.}~\bibnamefont {Bluestone}}, \bibinfo {author} {\bibfnamefont {N.}~\bibnamefont {Volet}}, \bibinfo {author} {\bibfnamefont {T.}~\bibnamefont {Komljenovic}}, \bibinfo {author} {\bibfnamefont {S.}~\bibnamefont {Hoon}}, \bibinfo {author} {\bibfnamefont {D.}~\bibnamefont {Yoon}}, \bibinfo {author} {\bibfnamefont {M.-G.}\ \bibnamefont {Suh}}, \bibinfo {author} {\bibfnamefont {K.}~\bibnamefont {Youl}}, \bibinfo
  {author} {\bibfnamefont {M.}~\bibnamefont {Pfeiffer}}, \bibinfo {author} {\bibfnamefont {T.}~\bibnamefont {Kippenberg}}, \bibinfo {author} {\bibfnamefont {E.}~\bibnamefont {Norberg}}, \bibinfo {author} {\bibfnamefont {K.}~\bibnamefont {Vahala}}, \bibinfo {author} {\bibfnamefont {K.}~\bibnamefont {Srinivasan}}, \bibinfo {author} {\bibfnamefont {N.}~\bibnamefont {Newbury}}, \bibinfo {author} {\bibfnamefont {L.}~\bibnamefont {Theogarajan}}, \bibinfo {author} {\bibfnamefont {J.}~\bibnamefont {Bowers}}, \bibinfo {author} {\bibfnamefont {S.}~\bibnamefont {Diddams}}, \ and\ \bibinfo {author} {\bibfnamefont {S.}~\bibnamefont {Papp}},\ }\href@noop {} {\bibfield  {journal} {\bibinfo  {journal} {Nature}\ }\textbf {\bibinfo {volume} {557}},\ \bibinfo {pages} {81} (\bibinfo {year} {2018})}\BibitemShut {NoStop}%
\bibitem [{\citenamefont {Kippenberg}, \citenamefont {Holzwarth},\ and\ \citenamefont {Diddams}(2011)}]{kippcomb}%
  \BibitemOpen
  \bibfield  {author} {\bibinfo {author} {\bibfnamefont {T.~J.}\ \bibnamefont {Kippenberg}}, \bibinfo {author} {\bibfnamefont {R.}~\bibnamefont {Holzwarth}}, \ and\ \bibinfo {author} {\bibfnamefont {S.~A.}\ \bibnamefont {Diddams}},\ }\href@noop {} {\bibfield  {journal} {\bibinfo  {journal} {Science}\ }\textbf {\bibinfo {volume} {332}},\ \bibinfo {pages} {555} (\bibinfo {year} {2011})}\BibitemShut {NoStop}%
\bibitem [{\citenamefont {Wang}\ \emph {et~al.}(2020)\citenamefont {Wang}, \citenamefont {Sciarrino}, \citenamefont {Laing},\ and\ \citenamefont {Thompson}}]{wang_integrated_2020}%
  \BibitemOpen
  \bibfield  {author} {\bibinfo {author} {\bibfnamefont {J.}~\bibnamefont {Wang}}, \bibinfo {author} {\bibfnamefont {F.}~\bibnamefont {Sciarrino}}, \bibinfo {author} {\bibfnamefont {A.}~\bibnamefont {Laing}}, \ and\ \bibinfo {author} {\bibfnamefont {M.~G.}\ \bibnamefont {Thompson}},\ }\href {\doibase 10.1038/s41566-019-0532-1} {\bibfield  {journal} {\bibinfo  {journal} {Nat. Photonics}\ }\textbf {\bibinfo {volume} {14}},\ \bibinfo {pages} {273} (\bibinfo {year} {2020})}\BibitemShut {NoStop}%
\bibitem [{\citenamefont {Liu}\ \emph {et~al.}(2021)\citenamefont {Liu}, \citenamefont {Huang}, \citenamefont {Wang}, \citenamefont {He}, \citenamefont {Raja}, \citenamefont {Liu}, \citenamefont {Engelsen},\ and\ \citenamefont {Kippenberg}}]{liu_high-yield_2021}%
  \BibitemOpen
  \bibfield  {author} {\bibinfo {author} {\bibfnamefont {J.}~\bibnamefont {Liu}}, \bibinfo {author} {\bibfnamefont {G.}~\bibnamefont {Huang}}, \bibinfo {author} {\bibfnamefont {R.~N.}\ \bibnamefont {Wang}}, \bibinfo {author} {\bibfnamefont {J.}~\bibnamefont {He}}, \bibinfo {author} {\bibfnamefont {A.~S.}\ \bibnamefont {Raja}}, \bibinfo {author} {\bibfnamefont {T.}~\bibnamefont {Liu}}, \bibinfo {author} {\bibfnamefont {N.~J.}\ \bibnamefont {Engelsen}}, \ and\ \bibinfo {author} {\bibfnamefont {T.~J.}\ \bibnamefont {Kippenberg}},\ }\href@noop {} {\bibfield  {journal} {\bibinfo  {journal} {Nat. Communications}\ }\textbf {\bibinfo {volume} {12}},\ \bibinfo {pages} {2236} (\bibinfo {year} {2021})}\BibitemShut {NoStop}%
\bibitem [{\citenamefont {Zhou}\ \emph {et~al.}(2023)\citenamefont {Zhou}, \citenamefont {Ou}, \citenamefont {Fang}, \citenamefont {Alkhazraji}, \citenamefont {Xu}, \citenamefont {Wan},\ and\ \citenamefont {Bowers}}]{zhou_prospects_2023}%
  \BibitemOpen
  \bibfield  {author} {\bibinfo {author} {\bibfnamefont {Z.}~\bibnamefont {Zhou}}, \bibinfo {author} {\bibfnamefont {X.}~\bibnamefont {Ou}}, \bibinfo {author} {\bibfnamefont {Y.}~\bibnamefont {Fang}}, \bibinfo {author} {\bibfnamefont {E.}~\bibnamefont {Alkhazraji}}, \bibinfo {author} {\bibfnamefont {R.}~\bibnamefont {Xu}}, \bibinfo {author} {\bibfnamefont {Y.}~\bibnamefont {Wan}}, \ and\ \bibinfo {author} {\bibfnamefont {J.~E.}\ \bibnamefont {Bowers}},\ }\href@noop {} {\bibfield  {journal} {\bibinfo  {journal} {eLight}\ }\textbf {\bibinfo {volume} {3}},\ \bibinfo {pages} {1} (\bibinfo {year} {2023})}\BibitemShut {NoStop}%
\bibitem [{\citenamefont {Bogaerts}\ \emph {et~al.}(2012)\citenamefont {Bogaerts}, \citenamefont {De~Heyn}, \citenamefont {Van~Vaerenbergh}, \citenamefont {De~Vos}, \citenamefont {Kumar~Selvaraja}, \citenamefont {Claes}, \citenamefont {Dumon}, \citenamefont {Bienstman}, \citenamefont {Van~Thourhout},\ and\ \citenamefont {Baets}}]{Siliconphotonicsring}%
  \BibitemOpen
  \bibfield  {author} {\bibinfo {author} {\bibfnamefont {W.}~\bibnamefont {Bogaerts}}, \bibinfo {author} {\bibfnamefont {P.}~\bibnamefont {De~Heyn}}, \bibinfo {author} {\bibfnamefont {T.}~\bibnamefont {Van~Vaerenbergh}}, \bibinfo {author} {\bibfnamefont {K.}~\bibnamefont {De~Vos}}, \bibinfo {author} {\bibfnamefont {S.}~\bibnamefont {Kumar~Selvaraja}}, \bibinfo {author} {\bibfnamefont {T.}~\bibnamefont {Claes}}, \bibinfo {author} {\bibfnamefont {P.}~\bibnamefont {Dumon}}, \bibinfo {author} {\bibfnamefont {P.}~\bibnamefont {Bienstman}}, \bibinfo {author} {\bibfnamefont {D.}~\bibnamefont {Van~Thourhout}}, \ and\ \bibinfo {author} {\bibfnamefont {R.}~\bibnamefont {Baets}},\ }\href@noop {} {\bibfield  {journal} {\bibinfo  {journal} {Laser Photonics Rev.}\ }\textbf {\bibinfo {volume} {6}},\ \bibinfo {pages} {47} (\bibinfo {year} {2012})}\BibitemShut {NoStop}%
\bibitem [{\citenamefont {Zhang}, \citenamefont {Böhm},\ and\ \citenamefont {Belkin}(2022)}]{zhang_mid-infrared_2022}%
  \BibitemOpen
  \bibfield  {author} {\bibinfo {author} {\bibfnamefont {K.}~\bibnamefont {Zhang}}, \bibinfo {author} {\bibfnamefont {G.}~\bibnamefont {Böhm}}, \ and\ \bibinfo {author} {\bibfnamefont {M.~A.}\ \bibnamefont {Belkin}},\ }\href@noop {} {\bibfield  {journal} {\bibinfo  {journal} {Appl. Phys. Lett}\ }\textbf {\bibinfo {volume} {120}},\ \bibinfo {pages} {061106} (\bibinfo {year} {2022})}\BibitemShut {NoStop}%
\bibitem [{\citenamefont {Pfeiffer}\ \emph {et~al.}(2016)\citenamefont {Pfeiffer}, \citenamefont {Kordts}, \citenamefont {Brasch}, \citenamefont {Zervas}, \citenamefont {Geiselmann}, \citenamefont {Jost},\ and\ \citenamefont {Kippenberg}}]{Pfeiffer:16}%
  \BibitemOpen
  \bibfield  {author} {\bibinfo {author} {\bibfnamefont {M.~H.~P.}\ \bibnamefont {Pfeiffer}}, \bibinfo {author} {\bibfnamefont {A.}~\bibnamefont {Kordts}}, \bibinfo {author} {\bibfnamefont {V.}~\bibnamefont {Brasch}}, \bibinfo {author} {\bibfnamefont {M.}~\bibnamefont {Zervas}}, \bibinfo {author} {\bibfnamefont {M.}~\bibnamefont {Geiselmann}}, \bibinfo {author} {\bibfnamefont {J.~D.}\ \bibnamefont {Jost}}, \ and\ \bibinfo {author} {\bibfnamefont {T.~J.}\ \bibnamefont {Kippenberg}},\ }\href@noop {} {\bibfield  {journal} {\bibinfo  {journal} {Optica}\ }\textbf {\bibinfo {volume} {3}},\ \bibinfo {pages} {20} (\bibinfo {year} {2016})}\BibitemShut {NoStop}%
\bibitem [{\citenamefont {Zhu}\ \emph {et~al.}(2021)\citenamefont {Zhu}, \citenamefont {Shao}, \citenamefont {Yu}, \citenamefont {Cheng}, \citenamefont {Desiatov}, \citenamefont {Xin}, \citenamefont {Hu}, \citenamefont {Holzgrafe}, \citenamefont {Ghosh}, \citenamefont {Shams-Ansari}, \citenamefont {Puma}, \citenamefont {Sinclair}, \citenamefont {Reimer}, \citenamefont {Zhang},\ and\ \citenamefont {Lon\v{c}ar}}]{Zhu:21}%
  \BibitemOpen
  \bibfield  {author} {\bibinfo {author} {\bibfnamefont {D.}~\bibnamefont {Zhu}}, \bibinfo {author} {\bibfnamefont {L.}~\bibnamefont {Shao}}, \bibinfo {author} {\bibfnamefont {M.}~\bibnamefont {Yu}}, \bibinfo {author} {\bibfnamefont {R.}~\bibnamefont {Cheng}}, \bibinfo {author} {\bibfnamefont {B.}~\bibnamefont {Desiatov}}, \bibinfo {author} {\bibfnamefont {C.~J.}\ \bibnamefont {Xin}}, \bibinfo {author} {\bibfnamefont {Y.}~\bibnamefont {Hu}}, \bibinfo {author} {\bibfnamefont {J.}~\bibnamefont {Holzgrafe}}, \bibinfo {author} {\bibfnamefont {S.}~\bibnamefont {Ghosh}}, \bibinfo {author} {\bibfnamefont {A.}~\bibnamefont {Shams-Ansari}}, \bibinfo {author} {\bibfnamefont {E.}~\bibnamefont {Puma}}, \bibinfo {author} {\bibfnamefont {N.}~\bibnamefont {Sinclair}}, \bibinfo {author} {\bibfnamefont {C.}~\bibnamefont {Reimer}}, \bibinfo {author} {\bibfnamefont {M.}~\bibnamefont {Zhang}}, \ and\ \bibinfo {author} {\bibfnamefont {M.}~\bibnamefont {Lon\v{c}ar}},\ }\href@noop {} {\bibfield  {journal} {\bibinfo  {journal} {Adv.
  Opt. Photon.}\ }\textbf {\bibinfo {volume} {13}},\ \bibinfo {pages} {242} (\bibinfo {year} {2021})}\BibitemShut {NoStop}%
\bibitem [{\citenamefont {Eggleton}, \citenamefont {Luther-Davies},\ and\ \citenamefont {Richardson}(2011)}]{eggleton2011chalcogenide}%
  \BibitemOpen
  \bibfield  {author} {\bibinfo {author} {\bibfnamefont {B.~J.}\ \bibnamefont {Eggleton}}, \bibinfo {author} {\bibfnamefont {B.}~\bibnamefont {Luther-Davies}}, \ and\ \bibinfo {author} {\bibfnamefont {K.}~\bibnamefont {Richardson}},\ }\href@noop {} {\bibfield  {journal} {\bibinfo  {journal} {Nat. Photonics}\ }\textbf {\bibinfo {volume} {5}},\ \bibinfo {pages} {141} (\bibinfo {year} {2011})}\BibitemShut {NoStop}%
\bibitem [{\citenamefont {Grayson}\ \emph {et~al.}(2022)\citenamefont {Grayson}, \citenamefont {Xu}, \citenamefont {Shanavas}, \citenamefont {Zohrabi}, \citenamefont {Bae}, \citenamefont {Gopinath},\ and\ \citenamefont {Park}}]{grayson2022fabrication}%
  \BibitemOpen
  \bibfield  {author} {\bibinfo {author} {\bibfnamefont {M.}~\bibnamefont {Grayson}}, \bibinfo {author} {\bibfnamefont {B.}~\bibnamefont {Xu}}, \bibinfo {author} {\bibfnamefont {T.}~\bibnamefont {Shanavas}}, \bibinfo {author} {\bibfnamefont {M.}~\bibnamefont {Zohrabi}}, \bibinfo {author} {\bibfnamefont {K.}~\bibnamefont {Bae}}, \bibinfo {author} {\bibfnamefont {J.~T.}\ \bibnamefont {Gopinath}}, \ and\ \bibinfo {author} {\bibfnamefont {W.}~\bibnamefont {Park}},\ }\href@noop {} {\bibfield  {journal} {\bibinfo  {journal} {Opt. Express}\ }\textbf {\bibinfo {volume} {30}},\ \bibinfo {pages} {31107} (\bibinfo {year} {2022})}\BibitemShut {NoStop}%
\bibitem [{\citenamefont {Kabakova}\ \emph {et~al.}(2013)\citenamefont {Kabakova}, \citenamefont {Pant}, \citenamefont {Choi}, \citenamefont {Debbarma}, \citenamefont {Luther-Davies}, \citenamefont {Madden},\ and\ \citenamefont {Eggleton}}]{Kabakova:13}%
  \BibitemOpen
  \bibfield  {author} {\bibinfo {author} {\bibfnamefont {I.~V.}\ \bibnamefont {Kabakova}}, \bibinfo {author} {\bibfnamefont {R.}~\bibnamefont {Pant}}, \bibinfo {author} {\bibfnamefont {D.-Y.}\ \bibnamefont {Choi}}, \bibinfo {author} {\bibfnamefont {S.}~\bibnamefont {Debbarma}}, \bibinfo {author} {\bibfnamefont {B.}~\bibnamefont {Luther-Davies}}, \bibinfo {author} {\bibfnamefont {S.~J.}\ \bibnamefont {Madden}}, \ and\ \bibinfo {author} {\bibfnamefont {B.~J.}\ \bibnamefont {Eggleton}},\ }\href {\doibase 10.1364/OL.38.003208} {\bibfield  {journal} {\bibinfo  {journal} {Opt. Lett.}\ }\textbf {\bibinfo {volume} {38}},\ \bibinfo {pages} {3208} (\bibinfo {year} {2013})}\BibitemShut {NoStop}%
\bibitem [{\citenamefont {Ko}\ \emph {et~al.}(2025)\citenamefont {Ko}, \citenamefont {Suk}, \citenamefont {Kim}, \citenamefont {Park}, \citenamefont {Sen}, \citenamefont {Kim}, \citenamefont {Wang}, \citenamefont {Dai}, \citenamefont {Wang}, \citenamefont {Wang} \emph {et~al.}}]{ko2025mid}%
  \BibitemOpen
  \bibfield  {author} {\bibinfo {author} {\bibfnamefont {K.}~\bibnamefont {Ko}}, \bibinfo {author} {\bibfnamefont {D.}~\bibnamefont {Suk}}, \bibinfo {author} {\bibfnamefont {D.}~\bibnamefont {Kim}}, \bibinfo {author} {\bibfnamefont {S.}~\bibnamefont {Park}}, \bibinfo {author} {\bibfnamefont {B.}~\bibnamefont {Sen}}, \bibinfo {author} {\bibfnamefont {D.-G.}\ \bibnamefont {Kim}}, \bibinfo {author} {\bibfnamefont {Y.}~\bibnamefont {Wang}}, \bibinfo {author} {\bibfnamefont {S.}~\bibnamefont {Dai}}, \bibinfo {author} {\bibfnamefont {X.}~\bibnamefont {Wang}}, \bibinfo {author} {\bibfnamefont {R.}~\bibnamefont {Wang}},  \emph {et~al.},\ }\href@noop {} {\bibfield  {journal} {\bibinfo  {journal} {Nat. Commun.}\ }\textbf {\bibinfo {volume} {16}},\ \bibinfo {pages} {2707} (\bibinfo {year} {2025})}\BibitemShut {NoStop}%
\bibitem [{\citenamefont {Li}\ \emph {et~al.}(2024)\citenamefont {Li}, \citenamefont {Xia}, \citenamefont {Cheng}, \citenamefont {Luo}, \citenamefont {Wang}, \citenamefont {Zeng}, \citenamefont {Yang}, \citenamefont {Li}, \citenamefont {Chen}, \citenamefont {Zhang},\ and\ \citenamefont {Li}}]{Li24Brillouin}%
  \BibitemOpen
  \bibfield  {author} {\bibinfo {author} {\bibfnamefont {Y.}~\bibnamefont {Li}}, \bibinfo {author} {\bibfnamefont {D.}~\bibnamefont {Xia}}, \bibinfo {author} {\bibfnamefont {H.}~\bibnamefont {Cheng}}, \bibinfo {author} {\bibfnamefont {L.}~\bibnamefont {Luo}}, \bibinfo {author} {\bibfnamefont {L.}~\bibnamefont {Wang}}, \bibinfo {author} {\bibfnamefont {S.}~\bibnamefont {Zeng}}, \bibinfo {author} {\bibfnamefont {S.}~\bibnamefont {Yang}}, \bibinfo {author} {\bibfnamefont {L.}~\bibnamefont {Li}}, \bibinfo {author} {\bibfnamefont {B.}~\bibnamefont {Chen}}, \bibinfo {author} {\bibfnamefont {B.}~\bibnamefont {Zhang}}, \ and\ \bibinfo {author} {\bibfnamefont {Z.}~\bibnamefont {Li}},\ }\href@noop {} {\bibfield  {journal} {\bibinfo  {journal} {Opt. Lett.}\ }\textbf {\bibinfo {volume} {49}},\ \bibinfo {pages} {4529} (\bibinfo {year} {2024})}\BibitemShut {NoStop}%
\bibitem [{\citenamefont {Grayson}\ \emph {et~al.}(2019)\citenamefont {Grayson}, \citenamefont {Zohrabi}, \citenamefont {Bae}, \citenamefont {Zhu}, \citenamefont {Gopinath},\ and\ \citenamefont {Park}}]{grayson2019enhancement}%
  \BibitemOpen
  \bibfield  {author} {\bibinfo {author} {\bibfnamefont {M.}~\bibnamefont {Grayson}}, \bibinfo {author} {\bibfnamefont {M.}~\bibnamefont {Zohrabi}}, \bibinfo {author} {\bibfnamefont {K.}~\bibnamefont {Bae}}, \bibinfo {author} {\bibfnamefont {J.}~\bibnamefont {Zhu}}, \bibinfo {author} {\bibfnamefont {J.~T.}\ \bibnamefont {Gopinath}}, \ and\ \bibinfo {author} {\bibfnamefont {W.}~\bibnamefont {Park}},\ }\href@noop {} {\bibfield  {journal} {\bibinfo  {journal} {Opt. Express}\ }\textbf {\bibinfo {volume} {27}},\ \bibinfo {pages} {33606} (\bibinfo {year} {2019})}\BibitemShut {NoStop}%
\bibitem [{\citenamefont {Xia}\ \emph {et~al.}(2023)\citenamefont {Xia}, \citenamefont {Yang}, \citenamefont {Zeng}, \citenamefont {Zhang}, \citenamefont {Wu}, \citenamefont {Wang}, \citenamefont {Zhao}, \citenamefont {Huang}, \citenamefont {Luo}, \citenamefont {Liu} \emph {et~al.}}]{xia2023integrated}%
  \BibitemOpen
  \bibfield  {author} {\bibinfo {author} {\bibfnamefont {D.}~\bibnamefont {Xia}}, \bibinfo {author} {\bibfnamefont {Z.}~\bibnamefont {Yang}}, \bibinfo {author} {\bibfnamefont {P.}~\bibnamefont {Zeng}}, \bibinfo {author} {\bibfnamefont {B.}~\bibnamefont {Zhang}}, \bibinfo {author} {\bibfnamefont {J.}~\bibnamefont {Wu}}, \bibinfo {author} {\bibfnamefont {Z.}~\bibnamefont {Wang}}, \bibinfo {author} {\bibfnamefont {J.}~\bibnamefont {Zhao}}, \bibinfo {author} {\bibfnamefont {J.}~\bibnamefont {Huang}}, \bibinfo {author} {\bibfnamefont {L.}~\bibnamefont {Luo}}, \bibinfo {author} {\bibfnamefont {D.}~\bibnamefont {Liu}},  \emph {et~al.},\ }\href@noop {} {\bibfield  {journal} {\bibinfo  {journal} {Lasers Photonics Rev.}\ }\textbf {\bibinfo {volume} {17}},\ \bibinfo {pages} {2200219} (\bibinfo {year} {2023})}\BibitemShut {NoStop}%
\bibitem [{\citenamefont {Shiryaev}\ \emph {et~al.}(2009)\citenamefont {Shiryaev}, \citenamefont {Troles}, \citenamefont {Houizot}, \citenamefont {Ketkova}, \citenamefont {Churbanov}, \citenamefont {Adam},\ and\ \citenamefont {Sibirkin}}]{shiryaev_preparation_2009}%
  \BibitemOpen
  \bibfield  {author} {\bibinfo {author} {\bibfnamefont {V.~S.}\ \bibnamefont {Shiryaev}}, \bibinfo {author} {\bibfnamefont {J.}~\bibnamefont {Troles}}, \bibinfo {author} {\bibfnamefont {P.}~\bibnamefont {Houizot}}, \bibinfo {author} {\bibfnamefont {L.~A.}\ \bibnamefont {Ketkova}}, \bibinfo {author} {\bibfnamefont {M.~F.}\ \bibnamefont {Churbanov}}, \bibinfo {author} {\bibfnamefont {J.~L.}\ \bibnamefont {Adam}}, \ and\ \bibinfo {author} {\bibfnamefont {A.~A.}\ \bibnamefont {Sibirkin}},\ }\href {\doibase 10.1016/j.optmat.2009.09.003} {\bibfield  {journal} {\bibinfo  {journal} {Opt. Mater.}\ }\textbf {\bibinfo {volume} {32}},\ \bibinfo {pages} {362} (\bibinfo {year} {2009})}\BibitemShut {NoStop}%
\bibitem [{\citenamefont {Zhang}\ \emph {et~al.}(2021)\citenamefont {Zhang}, \citenamefont {Zeng}, \citenamefont {Yang}, \citenamefont {Xia}, \citenamefont {Zhao}, \citenamefont {Sun}, \citenamefont {Huang}, \citenamefont {Song}, \citenamefont {Pan}, \citenamefont {Cheng}, \citenamefont {Choi},\ and\ \citenamefont {Li}}]{Zhang21As2S3}%
  \BibitemOpen
  \bibfield  {author} {\bibinfo {author} {\bibfnamefont {B.}~\bibnamefont {Zhang}}, \bibinfo {author} {\bibfnamefont {P.}~\bibnamefont {Zeng}}, \bibinfo {author} {\bibfnamefont {Z.}~\bibnamefont {Yang}}, \bibinfo {author} {\bibfnamefont {D.}~\bibnamefont {Xia}}, \bibinfo {author} {\bibfnamefont {J.}~\bibnamefont {Zhao}}, \bibinfo {author} {\bibfnamefont {Y.}~\bibnamefont {Sun}}, \bibinfo {author} {\bibfnamefont {Y.}~\bibnamefont {Huang}}, \bibinfo {author} {\bibfnamefont {J.}~\bibnamefont {Song}}, \bibinfo {author} {\bibfnamefont {J.}~\bibnamefont {Pan}}, \bibinfo {author} {\bibfnamefont {H.}~\bibnamefont {Cheng}}, \bibinfo {author} {\bibfnamefont {D.}~\bibnamefont {Choi}}, \ and\ \bibinfo {author} {\bibfnamefont {Z.}~\bibnamefont {Li}},\ }\href@noop {} {\bibfield  {journal} {\bibinfo  {journal} {Photon. Res.}\ }\textbf {\bibinfo {volume} {9}},\ \bibinfo {pages} {1272} (\bibinfo {year} {2021})}\BibitemShut {NoStop}%
\bibitem [{\citenamefont {Wang}\ \emph {et~al.}(2023)\citenamefont {Wang}, \citenamefont {Yang}, \citenamefont {Wang}, \citenamefont {Zhang}, \citenamefont {Wang},\ and\ \citenamefont {Xu}}]{wang2023high}%
  \BibitemOpen
  \bibfield  {author} {\bibinfo {author} {\bibfnamefont {Z.}~\bibnamefont {Wang}}, \bibinfo {author} {\bibfnamefont {Z.}~\bibnamefont {Yang}}, \bibinfo {author} {\bibfnamefont {H.}~\bibnamefont {Wang}}, \bibinfo {author} {\bibfnamefont {W.}~\bibnamefont {Zhang}}, \bibinfo {author} {\bibfnamefont {R.}~\bibnamefont {Wang}}, \ and\ \bibinfo {author} {\bibfnamefont {P.}~\bibnamefont {Xu}},\ }\href@noop {} {\bibfield  {journal} {\bibinfo  {journal} {Appl. Opt.}\ }\textbf {\bibinfo {volume} {62}},\ \bibinfo {pages} {2278} (\bibinfo {year} {2023})}\BibitemShut {NoStop}%
\bibitem [{\citenamefont {Zhang}\ \emph {et~al.}(2020)\citenamefont {Zhang}, \citenamefont {Jie}, \citenamefont {Zhang}, \citenamefont {Wang}, \citenamefont {Xie}, \citenamefont {Shi},\ and\ \citenamefont {Dai}}]{siliconracetrack}%
  \BibitemOpen
  \bibfield  {author} {\bibinfo {author} {\bibfnamefont {L.}~\bibnamefont {Zhang}}, \bibinfo {author} {\bibfnamefont {L.}~\bibnamefont {Jie}}, \bibinfo {author} {\bibfnamefont {M.}~\bibnamefont {Zhang}}, \bibinfo {author} {\bibfnamefont {Y.}~\bibnamefont {Wang}}, \bibinfo {author} {\bibfnamefont {Y.}~\bibnamefont {Xie}}, \bibinfo {author} {\bibfnamefont {Y.}~\bibnamefont {Shi}}, \ and\ \bibinfo {author} {\bibfnamefont {D.}~\bibnamefont {Dai}},\ }\href {\doibase 10.1364/PRJ.387816} {\bibfield  {journal} {\bibinfo  {journal} {Photon. Res.}\ }\textbf {\bibinfo {volume} {8}},\ \bibinfo {pages} {684} (\bibinfo {year} {2020})}\BibitemShut {NoStop}%
\bibitem [{\citenamefont {Kitoh}\ \emph {et~al.}(1995)\citenamefont {Kitoh}, \citenamefont {Takato}, \citenamefont {Yasu},\ and\ \citenamefont {Kawachi}}]{kitoh1992}%
  \BibitemOpen
  \bibfield  {author} {\bibinfo {author} {\bibfnamefont {T.}~\bibnamefont {Kitoh}}, \bibinfo {author} {\bibfnamefont {N.}~\bibnamefont {Takato}}, \bibinfo {author} {\bibfnamefont {M.}~\bibnamefont {Yasu}}, \ and\ \bibinfo {author} {\bibfnamefont {M.}~\bibnamefont {Kawachi}},\ }\href@noop {} {\bibfield  {journal} {\bibinfo  {journal} {J. Light. Technol.}\ }\textbf {\bibinfo {volume} {13}},\ \bibinfo {pages} {555} (\bibinfo {year} {1995})}\BibitemShut {NoStop}%
\bibitem [{\citenamefont {Kang}\ \emph {et~al.}(2025)\citenamefont {Kang}, \citenamefont {Jia}, \citenamefont {Yu}, \citenamefont {Gao}, \citenamefont {Bo}, \citenamefont {Zhang},\ and\ \citenamefont {Xu}}]{kang2025high}%
  \BibitemOpen
  \bibfield  {author} {\bibinfo {author} {\bibfnamefont {S.}~\bibnamefont {Kang}}, \bibinfo {author} {\bibfnamefont {D.}~\bibnamefont {Jia}}, \bibinfo {author} {\bibfnamefont {X.}~\bibnamefont {Yu}}, \bibinfo {author} {\bibfnamefont {F.}~\bibnamefont {Gao}}, \bibinfo {author} {\bibfnamefont {F.}~\bibnamefont {Bo}}, \bibinfo {author} {\bibfnamefont {G.}~\bibnamefont {Zhang}}, \ and\ \bibinfo {author} {\bibfnamefont {J.}~\bibnamefont {Xu}},\ }\href@noop {} {\bibfield  {journal} {\bibinfo  {journal} {Laser Photonics Rev.}\ }\textbf {\bibinfo {volume} {19}},\ \bibinfo {pages} {2400807} (\bibinfo {year} {2025})}\BibitemShut {NoStop}%
\bibitem [{\citenamefont {Zhu}\ \emph {et~al.}(2019)\citenamefont {Zhu}, \citenamefont {Zohrabi}, \citenamefont {Bae}, \citenamefont {Horning}, \citenamefont {Grayson}, \citenamefont {Park},\ and\ \citenamefont {Gopinath}}]{zhu2019nonlinear}%
  \BibitemOpen
  \bibfield  {author} {\bibinfo {author} {\bibfnamefont {J.}~\bibnamefont {Zhu}}, \bibinfo {author} {\bibfnamefont {M.}~\bibnamefont {Zohrabi}}, \bibinfo {author} {\bibfnamefont {K.}~\bibnamefont {Bae}}, \bibinfo {author} {\bibfnamefont {T.~M.}\ \bibnamefont {Horning}}, \bibinfo {author} {\bibfnamefont {M.~B.}\ \bibnamefont {Grayson}}, \bibinfo {author} {\bibfnamefont {W.}~\bibnamefont {Park}}, \ and\ \bibinfo {author} {\bibfnamefont {J.~T.}\ \bibnamefont {Gopinath}},\ }\href@noop {} {\bibfield  {journal} {\bibinfo  {journal} {Optica}\ }\textbf {\bibinfo {volume} {6}},\ \bibinfo {pages} {716} (\bibinfo {year} {2019})}\BibitemShut {NoStop}%
\bibitem [{\citenamefont {Jiang}, \citenamefont {Wu},\ and\ \citenamefont {Dai}(2018)}]{Jiang:18}%
  \BibitemOpen
  \bibfield  {author} {\bibinfo {author} {\bibfnamefont {X.}~\bibnamefont {Jiang}}, \bibinfo {author} {\bibfnamefont {H.}~\bibnamefont {Wu}}, \ and\ \bibinfo {author} {\bibfnamefont {D.}~\bibnamefont {Dai}},\ }\href {\doibase 10.1364/OE.26.017680} {\bibfield  {journal} {\bibinfo  {journal} {Opt. Express}\ }\textbf {\bibinfo {volume} {26}},\ \bibinfo {pages} {17680} (\bibinfo {year} {2018})}\BibitemShut {NoStop}%
\bibitem [{\citenamefont {Hu}\ \emph {et~al.}(2021)\citenamefont {Hu}, \citenamefont {Tian}, \citenamefont {Li}, \citenamefont {Zhang}, \citenamefont {Ren}, \citenamefont {Qi}, \citenamefont {Yang}, \citenamefont {Feng},\ and\ \citenamefont {Yang}}]{Hu:21}%
  \BibitemOpen
  \bibfield  {author} {\bibinfo {author} {\bibfnamefont {Y.}~\bibnamefont {Hu}}, \bibinfo {author} {\bibfnamefont {K.}~\bibnamefont {Tian}}, \bibinfo {author} {\bibfnamefont {T.}~\bibnamefont {Li}}, \bibinfo {author} {\bibfnamefont {M.}~\bibnamefont {Zhang}}, \bibinfo {author} {\bibfnamefont {H.}~\bibnamefont {Ren}}, \bibinfo {author} {\bibfnamefont {S.}~\bibnamefont {Qi}}, \bibinfo {author} {\bibfnamefont {A.}~\bibnamefont {Yang}}, \bibinfo {author} {\bibfnamefont {X.}~\bibnamefont {Feng}}, \ and\ \bibinfo {author} {\bibfnamefont {Z.}~\bibnamefont {Yang}},\ }\href {\doibase 10.1364/OME.412731} {\bibfield  {journal} {\bibinfo  {journal} {Opt. Mater. Express}\ }\textbf {\bibinfo {volume} {11}},\ \bibinfo {pages} {695} (\bibinfo {year} {2021})}\BibitemShut {NoStop}%
\bibitem [{\citenamefont {Choi}\ \emph {et~al.}(2016)\citenamefont {Choi}, \citenamefont {Han}, \citenamefont {Sohn}, \citenamefont {Chen}, \citenamefont {Smith}, \citenamefont {Kimerling}, \citenamefont {Richardson}, \citenamefont {Agarwal},\ and\ \citenamefont {Tan}}]{choi2016nonlinear}%
  \BibitemOpen
  \bibfield  {author} {\bibinfo {author} {\bibfnamefont {J.~W.}\ \bibnamefont {Choi}}, \bibinfo {author} {\bibfnamefont {Z.}~\bibnamefont {Han}}, \bibinfo {author} {\bibfnamefont {B.-U.}\ \bibnamefont {Sohn}}, \bibinfo {author} {\bibfnamefont {G.~F.}\ \bibnamefont {Chen}}, \bibinfo {author} {\bibfnamefont {C.}~\bibnamefont {Smith}}, \bibinfo {author} {\bibfnamefont {L.~C.}\ \bibnamefont {Kimerling}}, \bibinfo {author} {\bibfnamefont {K.~A.}\ \bibnamefont {Richardson}}, \bibinfo {author} {\bibfnamefont {A.~M.}\ \bibnamefont {Agarwal}}, \ and\ \bibinfo {author} {\bibfnamefont {D.~T.}\ \bibnamefont {Tan}},\ }\href@noop {} {\bibfield  {journal} {\bibinfo  {journal} {Sci. Rep.}\ }\textbf {\bibinfo {volume} {6}},\ \bibinfo {pages} {39234} (\bibinfo {year} {2016})}\BibitemShut {NoStop}%
\bibitem [{\citenamefont {Li}\ \emph {et~al.}(2022)\citenamefont {Li}, \citenamefont {Wang}, \citenamefont {Zhang}, \citenamefont {Chen}, \citenamefont {Lin}, \citenamefont {Dai},\ and\ \citenamefont {Ji}}]{li_study_2022}%
  \BibitemOpen
  \bibfield  {author} {\bibinfo {author} {\bibfnamefont {Z.}~\bibnamefont {Li}}, \bibinfo {author} {\bibfnamefont {K.}~\bibnamefont {Wang}}, \bibinfo {author} {\bibfnamefont {J.}~\bibnamefont {Zhang}}, \bibinfo {author} {\bibfnamefont {F.}~\bibnamefont {Chen}}, \bibinfo {author} {\bibfnamefont {C.}~\bibnamefont {Lin}}, \bibinfo {author} {\bibfnamefont {S.}~\bibnamefont {Dai}}, \ and\ \bibinfo {author} {\bibfnamefont {W.}~\bibnamefont {Ji}},\ }\href {\doibase https://doi.org/10.1016/j.jnoncrysol.2022.121628} {\bibfield  {journal} {\bibinfo  {journal} {J. Non-Cryst. Solids}\ }\textbf {\bibinfo {volume} {588}},\ \bibinfo {pages} {121628} (\bibinfo {year} {2022})}\BibitemShut {NoStop}%
\bibitem [{\citenamefont {Parnell}\ \emph {et~al.}(2019)\citenamefont {Parnell}, \citenamefont {Furniss}, \citenamefont {Tang}, \citenamefont {Fang}, \citenamefont {Benson}, \citenamefont {Canedy}, \citenamefont {Kim}, \citenamefont {Kim}, \citenamefont {Merritt}, \citenamefont {Bewley}, \citenamefont {Vurgaftman}, \citenamefont {Meyer},\ and\ \citenamefont {Seddon}}]{Parnell:19}%
  \BibitemOpen
  \bibfield  {author} {\bibinfo {author} {\bibfnamefont {H.}~\bibnamefont {Parnell}}, \bibinfo {author} {\bibfnamefont {D.}~\bibnamefont {Furniss}}, \bibinfo {author} {\bibfnamefont {Z.}~\bibnamefont {Tang}}, \bibinfo {author} {\bibfnamefont {Y.}~\bibnamefont {Fang}}, \bibinfo {author} {\bibfnamefont {T.~M.}\ \bibnamefont {Benson}}, \bibinfo {author} {\bibfnamefont {C.~L.}\ \bibnamefont {Canedy}}, \bibinfo {author} {\bibfnamefont {C.~S.}\ \bibnamefont {Kim}}, \bibinfo {author} {\bibfnamefont {M.}~\bibnamefont {Kim}}, \bibinfo {author} {\bibfnamefont {C.~D.}\ \bibnamefont {Merritt}}, \bibinfo {author} {\bibfnamefont {W.~W.}\ \bibnamefont {Bewley}}, \bibinfo {author} {\bibfnamefont {I.}~\bibnamefont {Vurgaftman}}, \bibinfo {author} {\bibfnamefont {J.~R.}\ \bibnamefont {Meyer}}, \ and\ \bibinfo {author} {\bibfnamefont {A.~B.}\ \bibnamefont {Seddon}},\ }\href {\doibase 10.1364/OME.9.003616} {\bibfield  {journal} {\bibinfo  {journal} {Opt. Mater. Express}\ }\textbf {\bibinfo {volume} {9}},\ \bibinfo {pages} {3616}
  (\bibinfo {year} {2019})}\BibitemShut {NoStop}%
\bibitem [{\citenamefont {Irfan}, \citenamefont {Kim},\ and\ \citenamefont {Kurt}(2024)}]{irfan_ultra-compact_2024}%
  \BibitemOpen
  \bibfield  {author} {\bibinfo {author} {\bibfnamefont {S.}~\bibnamefont {Irfan}}, \bibinfo {author} {\bibfnamefont {J.-Y.}\ \bibnamefont {Kim}}, \ and\ \bibinfo {author} {\bibfnamefont {H.}~\bibnamefont {Kurt}},\ }\href@noop {} {\bibfield  {journal} {\bibinfo  {journal} {Sci. Rep.}\ }\textbf {\bibinfo {volume} {14}},\ \bibinfo {pages} {6453} (\bibinfo {year} {2024})}\BibitemShut {NoStop}%
\end{thebibliography}%

\end{document}